\documentclass{ws-procs9x6}

\begin{document}
\title{Pentaquark resonances from collision times}

\author{N. G. Kelkar and M. Nowakowski} 

\address{Departamento de Fisica, Universidad de los Andes, \\
Cra. 1 No. 18A-10, Santafe de Bogota, Colombia\\ }

\maketitle

\abstracts{Having successfully explored the existing relations between the 
$S$-matrix and collision times in scattering reactions to study the 
conventional baryon and meson resonances, the method is now extended to
the exotic sector. To be specific, the  
collision time in various partial waves of $K^+ N$ elastic 
scattering is evaluated using phase shifts extracted from the $K^+ N 
\rightarrow K^+ N$ data as well as from model dependent $T$-matrix 
solutions. 
We find several pentaquark resonances including some low-lying ones
around 1.5 to 1.6 GeV in the $P_{01}$, $P_{03}$ and $D_{03}$ partial
waves of $K^+ N$ elastic scattering.}

\section{Introduction}
The discovery of the pion in 1947 followed by that of several other 
mesons and baryons, gave birth to a specialized branch in particle 
physics which involved the {\it characterization of hadronic resonances}. 
However, even after half a century's experience in analyzing experimental 
data to infer on the existence of resonances we still come across examples
where a resonance is confirmed by one type of analysis and is reported to
be absent by another and history shows that this is especially true in 
the case of the pentaquark ($Z^*$) resonances.
It is therefore important to examine the 
limitations of the various theoretical definitions used to extract 
information from data and then comment on the existence of the resonance.
The pentaquark resonance, $\Theta ^+(1540)$, found in several experiments\cite{allexp} 
which followed its theoretical prediction\cite{diakonov} is one such 
recent example. In the present talk
we try to shed some light on the controversy\cite{pras} of its existence 
using a somewhat forgotten 
but well-documented method of collision time or time delay in scattering.
In fact, we identify several pentaquark resonances by 
evaluating the time delay in various partial waves of $K^+ N$ elastic
scattering using the available $K^+ N \rightarrow K^+ N$ data. 

\section{Collision time: From the fifties until now}
Intuitively, one would expect that if a resonance is formed as an 
intermediate state in a scattering process 
(say $a + b \rightarrow R \rightarrow  a + b$), 
then the scattered particles in the final state would emerge  
(alone from the fact that the resonance has a finite lifetime) later 
than in a non-resonant process $a + b \rightarrow a + b$. The 
resonant process would be ``delayed" as compared to the non-resonant
one. This relevance of the delay time or collision time in 
scattering processes to 
resonance physics was noticed back in the fifties by Eisenbud\cite{eis}, 
Bohm\cite{boh} and Wigner\cite{wig}. Wigner considered a simple wave 
packet with the superposition of two frequencies 
to show that the amount of time by which an incoming 
particle in a 
scattering process got delayed due to interaction with the 
scattering centre is proportional to 
the energy derivative of the scattering phase shift, $\delta(E)$. 
For a wavepacket consisting of two terms with frequencies $\nu \pm \Delta \nu$, 
wave numbers $k \pm \Delta k$ and phase shifts $2(\delta \pm \Delta \delta)$,  
the incident and outgoing waves can be written as, 
$$\Psi_{inc} = 2v^{-1}[e^{i(kx+\nu t)}cos(x\Delta k + t\Delta \nu)]$$
and
$$\Psi_{out} = 2v^{-1}[e^{i(kx -\nu t + 2\delta)} cos(x\Delta k -t\Delta
\nu + 2\Delta \delta)]$$
where one can see that $\Psi_{inc}$ has a maximum when $x_{max} = -t(d\nu/dk) 
= - vt$, and represents a particle moving inwards at times $t < 0$ whereas
$\Psi_{out}$ has $x_{max} = vt - 2 d\delta/dk$ for a later time $t$. Thus
the interaction has {\it delayed} the particle by an amount 
\begin{equation}
\Delta t = 2 v^{-1} {d\delta \over dk} = 2 \hbar {d\delta \over dE}\,.
\end{equation}
In the absence of interaction, obviously, $\delta = 0$ and $d\delta/dk = 0$ 
and there is no time delay. From this, Wigner concluded that {\it close to
resonances, where the incident particle is captured and retained for 
some time by the scattering centre, $d\delta/dk$ will assume large 
positive values}. However, in the case of non-resonant scattering, the 
interaction can sometimes also speed up the scattering process resulting
in a negative time delay or time advancement. The negative time delay cannot
take arbitrarily large values and in fact Wigner put a limit from the 
principle of causality as,
$${d \delta \over dk} > - a$$
where $a$ is the radius of the scattering centre.

Eisenbud\cite{eis} defined a delay time 
matrix, $\mbox{\boldmath $\Delta$} {\bf t}$, in terms of the
scattering matrix {\bf S}, where a typical element of 
$\mbox{\boldmath $\Delta$} {\bf t}$, 
\begin{equation}
\Delta t_{ij} = Re \big [ -i \hbar (S_{ij})^{-1} {dS_{ij} \over dE}
\big ]\, ,
\label{eisen}
\end{equation}
gave the delay in the outgoing signal in the $j^{th}$ channel when 
the signal is injected in the $i^{th}$ channel. For an elastic scattering
reaction, $i=j$ and one can easily see that using a phase shift formulation
of the $S$-matrix, i.e. $S = e^{2i\delta}$ in the purely elastic case and 
$S = \eta e^{2i\delta}$ for elastic scattering in the presence of 
inelasticities, the above relation reduces simply to\cite{me1,we3we4}
\begin{equation}
\Delta t_{ii} = 2 \hbar {d\delta \over dE}\, .
\label{phshiftder}
\end{equation}
Henceforth for simplicity, we shall drop the subscripts $ii$ and 
write $\Delta t$ whenever we refer to time delay in elastic scattering. 
Since the particle
has probability $|S_{ij}|^2$ of emerging in the $j^{th}$ channel,
the average time delay for a particle injected in the $i^{th}$ channel
is given as,
\begin{eqnarray}\label{eisav}
\langle\, \Delta t_i \,\rangle_{av} &=& \sum_j \, S_{ij}^* S_{ij} \,
\Delta t_{ij} \, \nonumber \\
&=& \,Re \biggr[ \,-i\hbar \,\sum_j S^*_{ij} \,{dS_{ij} \over dE}
\biggr] \, .
\end{eqnarray}

Later on, Smith\cite{smi} constructed a lifetime matrix ${\bf Q}$, 
which was given in terms of the scattering matrix, ${\bf S}$ as,
\begin{equation}
{\bf Q} =  i\hbar\, {\bf S}\, {d{\bf S}^{\dag} \over dE} \,.
\end{equation} 
He defined collision time to be the limit as 
$R \to \infty $, of the difference between the time the particles spend 
within a distance $R$ of each other (with interaction) and the time
they would have spent there without interaction. The matrix elements
of ${\bf Q}$ (which is hermitian) were given by,
\begin{eqnarray}
Q_{ij} &=& \lim_{R \to \infty} [ \int^{r<R} \Psi_i \Psi_j^* d\tau - 
R(v_i^{-1} \delta_{ij} + \sum_k S_{ik} v_k^{-1} S_{jk}^*)]_{av} \nonumber \\
&=& i \hbar \sum_n S_{in} \biggl( {dS_{jn}^* \over dE} \biggr)
\end{eqnarray}
where $S_{ik}$ is an element of ${\bf S}$ and $Q_{ij}$ is finite 
if the interaction vanishes rapidly at large $R$.
One can now see that the average time delay for a collision beginning in 
the $i^{th}$ channel calculated using Eisenbud's 
$\mbox{\boldmath $\Delta$} {\bf t}$ (Eq. \ref{eisav} ) as 
above, is indeed the matrix 
element $Q_{ii}$ of the lifetime matrix. Smith concluded that {\it when 
$Q_{ii}$'s are positive and large, we have a criterion for the existence
of metastable states}. 

The interest in this concept continued in the sixties and Goldberger 
and Watson\cite{gol}, using the concept of time interval in $S$-matrix 
theory found that 
\begin{equation}
\Delta t = - i \hbar \,{d [ln S(E)] \over dE} \, .
\end{equation}
Lippmann\cite{lip} 
even defined a time delay operator,  
\begin{equation}
\mbox {\boldmath $\tau$} = - i \hbar \partial / \partial E \,, 
\end{equation}
the 
expectation value of which (using the phase shift formulation of the
$S$-matrix) gave the time delay to be the same as in Eq.~(\ref{phshiftder}). 

In the seventies, the time delay concept finally found a place 
in most books on scattering theory and quantum mechanics\cite{scatt},  
where it is mentioned as a necessary
condition for the existence of a resonance. However, in spite of being 
so well-known in literature as well as books, it was rarely used to
characterize resonances until its recent application\cite{me1,we3we4}
to meson and baryon resonances. 
Instead, mathematical definitions of 
a resonance have been used over the decades for its identification and
characterization. 
The simple physical concept of time delay was somehow always overlooked 
in practice. In what 
follows, we now analyse the shortcomings of the various definitions or 
tools used to locate resonances. 

\section{What is a resonance?}
A resonance is theoretically clearly defined as an unstable state 
characterized by different quantum numbers. However, to identify such 
a state when it has been produced, one needs to define a resonance in 
terms of theoretical quantities which can be extracted from data. In 
principle, if an unstable state is formed for example in a scattering 
process, then the various definitions should simply serve as complementary 
tools for its confirmation. However, it does often happen that a resonance 
extracted using one definition appears to be ``missing" within another. 
Before discarding the existence of such missing resonances, it is 
important to take into account the limitations of the various definitions
of a resonance. We shall discuss these below.
\subsection{S-matrix poles}
The most conventional method of locating a resonance involves 
assuming that whenever an unstable particle is formed, 
there exists a corresponding pole of the $S$-matrix on the unphysical 
sheet of the complex energy plane lying close to the real axis\cite{scatt}.
The experimental data is usually fitted with a model dependent 
$S$-matrix and resonances are identified by locating the poles. 
However, Calucci and co-workers\cite{cal1cal2} 
took a different point of view. 
In the case of a resonance $R$ formed in a two body elastic scattering 
process, $a + b \rightarrow R \rightarrow a + b$, a sharp peak in 
the cross section 
accompanied by a rapid variation of the phase shift through $\pi /2$ 
with positive derivative (essentially the condition for large positive 
time delay) was taken as the signal for the 
existence of a resonance. The authors then constructed $S$-matrices 
satisfying all requirements of analyticity, unitarity and threshold and
asymptotic behaviour in energy such that a sharp isolated resonance 
is produced without an accompanying pole on the unphysical sheet. 
They also ensured the exponential decay of such a state. It is both 
interesting and relevant to note that while concluding 
that resonances can belong to a ``no-pole category"\cite{fon}, 
the authors stressed the need for high accuracy data in the case of the
$Z^*$'s (the pentaquark resonances) whose dynamical origin might be
questionable.

\subsection{Cross section bumps, Argand diagrams and Speed Plots}
Though the existence of a resonance usually produces a large bump in
the cross sections, it was shown in a pedagogic article by 
Ohanian and Ginsburg\cite{oha} that a maximum of the scattering
probability (i.e. cross section) cannot be taken as a sufficient
condition for the existence of a resonance. 
Resonances can also be identified from anticlockwise loops in the
Argand diagrams of the complex scattering amplitude; however, these
alone cannot gaurantee the existence of a resonance\cite{mascol}.
Finally, the speed plot peaks, i.e. peaks in 
\begin{equation}
SP(E) = \biggl|{dT \over dE}\biggr|\,,
\end{equation} 
where $T$ is the complex scattering transition matrix, can in fact be 
ambiguous due to being positive definite by definition\cite{me1}.

Given the ambiguities associated with each of the techniques used to
identify resonances, they should rather be used as complementary tools. 
We shall present the results of a time delay 
analysis of the $K^+ N$ elastic scattering using the
existing $K^+ N \rightarrow K^+ N$ data as well as the SP92\cite{arndt} 
model dependent $T$-matrix solutions. However, before going over to the 
characterization of the $Z^*$ resonances
through the time delay analysis, we give a brief review of the history
of the identification of these pentaquark resonances. 
We will see that the old determinations of the pentaquarks were just as
controversial as the most recently discovered $\Theta^+$.   

\section{Historical evidences of the $Z^*$'s}
\subsection{Earliest evidences}
The search for the strangeness, $S = +1$ exotic baryons 
started back in the late fifties when 
Burrowes {\it et al.}\cite{burrowes} measured the Kaon-Nucleon total cross 
sections from 0.6 to 2 GeV centre of mass energy with the hope that the 
total cross sections might exhibit a resonance analogous to the pion-nucleon
behaviour. Indeed, they found a peak in the total $K^+N$ cross sections
around 1.8 GeV centre of mass energy. In 1966, Cool {\it et al.}\cite{cool} 
reported measurements of the $K^+ p$ and $K^+ d$ total cross sections with 
increased precision and a possible $Z^*$ with mass $M \sim 1.9$ GeV and 
width $\Gamma \sim 180$ MeV. These were followed by searches for the exotic
baryons in photoproduction experiments. Tyson {\it et al.} performed 
experiments\cite{tyson} for the photoproduction of negative K mesons 
where the excitation function for the
$K^-$ yield in the reaction $\gamma + p \rightarrow K^- + Z^*$ was 
fitted by a three resonance formula in the missing mass range 
1500 - 2500 MeV. The best fit masses and widths for the three resonances
fitted were as follows: $M = 1860^{+60}_{-20}$, $\Gamma \sim 150$ MeV; 
$M = 2125 \pm 25$ MeV, $\Gamma \sim 75$ MeV and $M = 2280 \pm 20$ MeV, 
$\Gamma \sim 80$ MeV. With the availability of the $K^+p$ data, 
energy dependent phase shift analyses were performed\cite{leamartin} and
a resonance in the $P_{11}$ partial wave of $K^+p$ elastic scattering 
was reported (on the basis of Argand diagrams) 
around $M \sim 2$ GeV and $\Gamma \sim 220$ MeV. 

\subsection{Seventies and eighties}
The seventies mostly saw the confirmation of the $Z^*$'s through several
partial wave analyses. S. Kato {\it et al.}\cite{kato} obtained four possible
solutions from a phase shift analysis of $K^+p$ elastic scattering 
and on the basis of Argand diagrams concluded on a possible mass of 
the $Z^*$ around 1.9 to 2 GeV
and $\Gamma \sim 130 - 250$ MeV in the $P_{13}$ partial wave. This work
was followed by two articles by Aaron {\it et al.}\cite{aaron} which reported
evidence for the $Z^*$'s in the $D_{03}$, $S_{01}$ and $P_{01}$ partial
waves using Argand diagrams and speed plots. Arndt {\it et al.}\cite{arndt78} 
fitted 
the $K^+ p$ data with a coupled-channel $K$-matrix parametrization 
and found a $P_{13}$ resonance pole at ($1.796 - i101$) GeV. 
With so much support gathering in favour of the existence of these
exotic baryons, the $Z^*$'s finally found a place in the Particle 
Data Group Compilation in 1982. 
In 1984, Keiji Hashimoto\cite{hashimoto} performed a single-energy phase 
shift analysis of the
data then available and reported $Z^*$ resonances in the $P_{13}$, 
$D_{03}$, $P_{11}$ and $D_{05}$ partial waves of $K^+$-nucleon scattering.

In 1982 appeared yet another partial wave analysis 
of the $K^+$-nucleon scattering data by Nakajima {\it et al}\cite{nakajima}. 
On the basis 
of the counterclockwise motion in Argand diagrams, they reported three
resonances: $M = 1.931$ GeV, $\Gamma = 347$ MeV in $P_{13}$;      
$M = 1.778$ GeV, $\Gamma = 662$ MeV in $P_{01}$ and       
$M = 1.907$ GeV, $\Gamma = 291$ MeV in the $D_{03}$ partial wave. {\it A 
prominent bump in the speed plot of the $P_{01}$ partial wave 
at $1540$ MeV was however
ignored and not mentioned as a resonance due to lack of support from
the Argand diagram}. 

In the late eighties and nineties, unfortunately, a general reluctance 
to accept the $Z^*$'s as genuine resonances or unstable pentaquark 
states started building up. Articles appeared in literature where the 
authors were often too careful and labeled the resonances in different
ways as `doorways', `pseudoresonances', `resonance-like structures',
`complicated structures in the unphysical sheet' etc. The main reason 
for such labeling was that the criteria applied to establish the $Z^*$'s  
were too stringent; sometimes more stringent than those applied for the
conventional baryon resonances. Hence 1992 saw the last appearance of
the $Z^*$ in the Particle Data Group Compilation with the remark, ``It 
might take 20 years before the issue of the existence of the $Z^*$ 
resonances is settled". 

\subsection{The year 2003}   
The interest in the pentaquark states was greatly revived\cite{allothers} 
by the discovery
of a narrow exotic state by different experimental groups around a mass
of 1540 MeV\cite{allexp}. The motivation for the first experiments came
from a theoretical prediction\cite{diakonov} and this low lying $Z^*$
was renamed as $\Theta^+$. Before closing this section on the history
of the $Z^*$'s, it is worth mentioning that just a little before the
above experiments reported the $\Theta^+$, a time delay analysis\cite{weKN} 
of the
old $K^+ N$ scattering data confirmed several penatquark states 
in the 1.8 GeV region 
and revealed some new pentaquark states around $1.5 - 1.6$ GeV in the
$P_{01}$, $P_{13}$ and $D_{03}$ partial waves. In the next section, we 
shall discuss the results of this analysis.

A good account of the history of the theoretical progress in the 
search of the exotics (mesons as well as baryons) can also be found in an
article by D. P. Roy\cite{roydp}.
 
\section{Time delay in $K^+ N$ elastic scattering}
\subsection{Energy dependent calculations}
We shall first present the time delay distributions (as a function of
energy) using model dependent solutions of the $T$-matrix. 
Replacing $S = 1 + 2iT$ in Eq.~(\ref{eisen}), the time delay in 
elastic scattering, in terms of the $T$-matrix\cite{weKN} can be 
obtained from: 
\begin{equation}
S^*_{ii} S_{ii}\, \Delta t_{ii} = 2 \hbar \biggl[ {dT^R_{ii} \over dE}
+ 2 T^R_{ii}
{dT^I_{ii} \over dE} - 2 T^I_{ii} {dT^R_{ii} \over dE} \biggr]
\end{equation}
where $T$ contains the information of resonant and non-resonant scattering 
and is complex ($T = T^R + iT^I$). 
As can be seen in Fig. 1, 
in addition to the resonances around 1.8 GeV, we find some low-lying 
ones around 1.5-1.6 GeV. Table I shows that the time delay peak positions  
around 1.8 GeV agree with the pole positions obtained from the same 
$T$-matrix. 
However, the low-lying ones do not correspond to any poles.
These peaks could possibly be considered as realistic 
examples of the no-pole category of resonances\cite{fon} 
mentioned in the previous section.
However, it cannot be doubted that the resonances around $1.5$ GeV found
using time delay have something in common with the recently found
peaks in the experimental cross sections around $1.54$ GeV.
At this point we note again that a speed plot 
peak at 1.54 GeV in the $P_{01}$ partial wave of $K^+N$ elastic scattering 
was already noted by Nakajima {\it et al.}\cite{nakajima}. 
However, due to lack of support from Argand
diagrams they did not mention it as a pentaquark resonance.
\begin{figure}[h]
\epsfxsize=4cm 
\centerline{\epsfxsize=3.1in\epsfbox{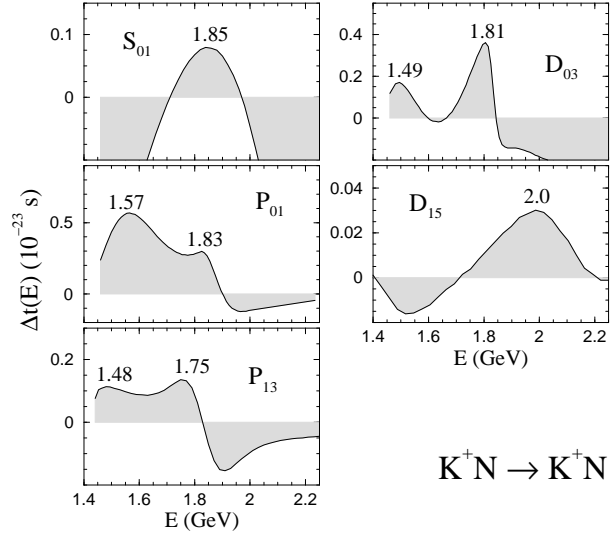}}   
\caption{Time delay in various partial waves of $K^+ N \rightarrow K^+ N$ 
elastic scattering, evaluated from the SP92 $T$-matrix solutions.}
\end{figure}
\begin{table}[h]
\tbl{Comparison of time delay peaks with pole values}
{\begin{tabular}{@{}ccc@{}}
\hline
\multicolumn{3}{c}{}\\[2ex]
Partial wave & SP92 pole position (GeV) & Position of time delay peak \\
\hline
\multicolumn{3}{c}{}\\[2ex]
$S_{01}$ & - & {1.85}\\
 & & \\
$P_{01}$ & - & {1.57}\\
         & {1.831} - $i$95 & {1.83} \\
 & & \\
$P_{13}$ & - & {1.48} \\
         & {1.811} - $i$118 & {1.75}\\
 & & \\
$D_{03}$ & - & {1.49} \\
         & {1.788} - $i$170 & {1.81} \\
 & & \\
$D_{15}$ & {2.074} - $i$253 & {2.0} \\
\hline
\end{tabular}}
\end{table}
    
\subsection{Pentaquark resonances from single energy values of 
$K^+N$ phase shifts}
\begin{figure}
\centerline{\epsfxsize=1.9in\epsfbox{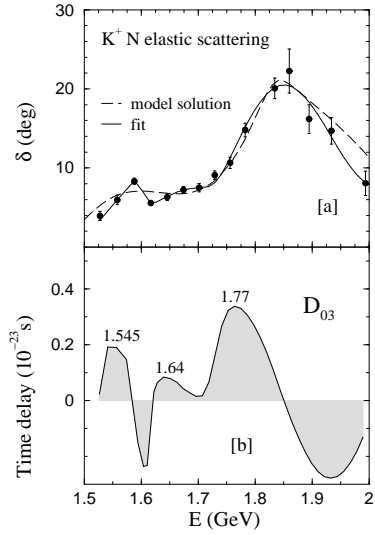}}   
\caption{Time delay in the $D_{03}$ partial wave of 
$K^+ N \rightarrow K^+ N$ elastic scattering, evaluated from a fit (solid
line in [a]) to the single energy values of the phase shift.}
\end{figure}
\begin{figure}
\centerline{\epsfxsize=1.9in\epsfbox{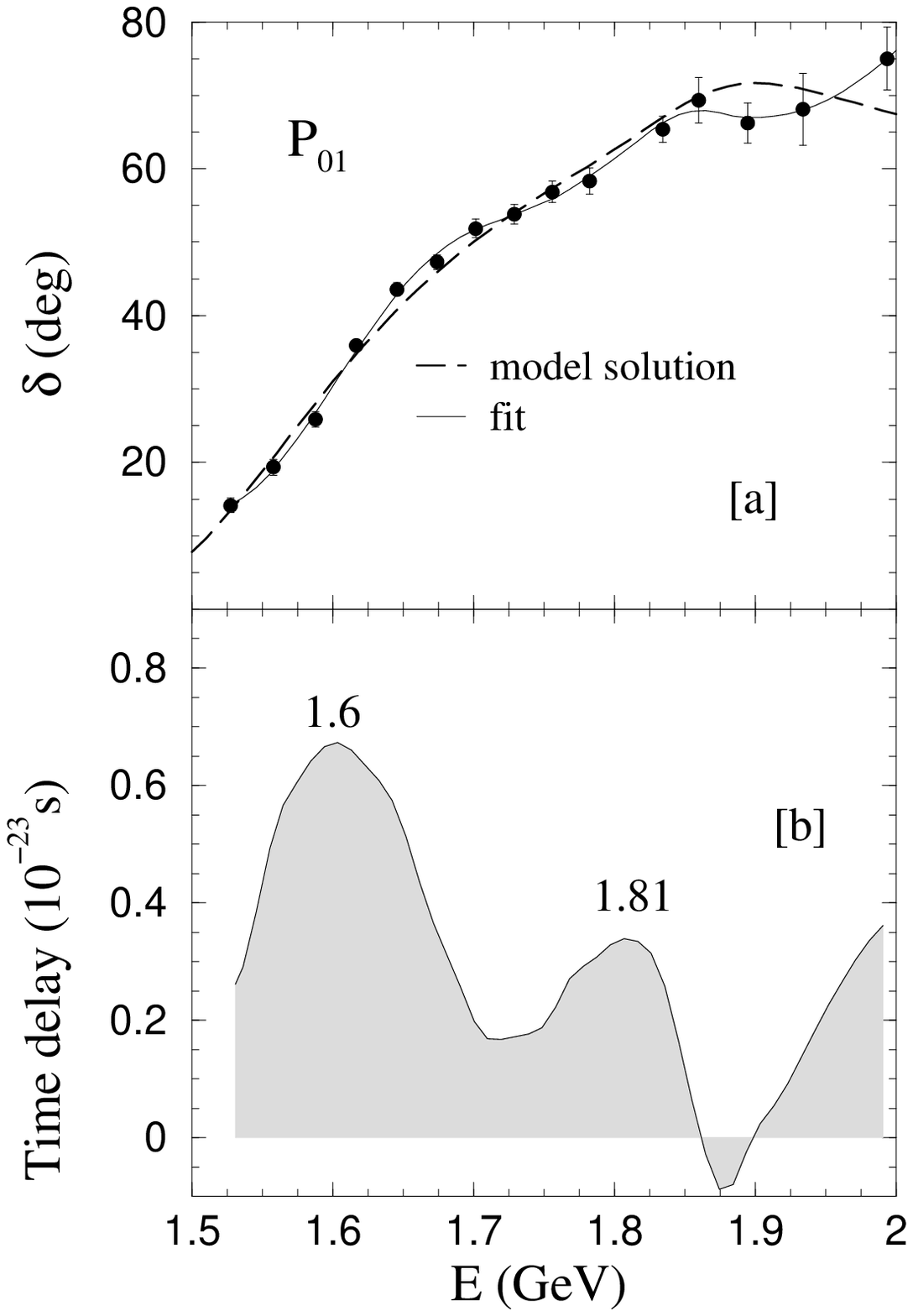}} 
\caption{Same as Fig. 2, but for the $P_{01}$ 
partial wave.}
\end{figure}
\begin{figure}
\centerline{\epsfxsize=1.9in\epsfbox{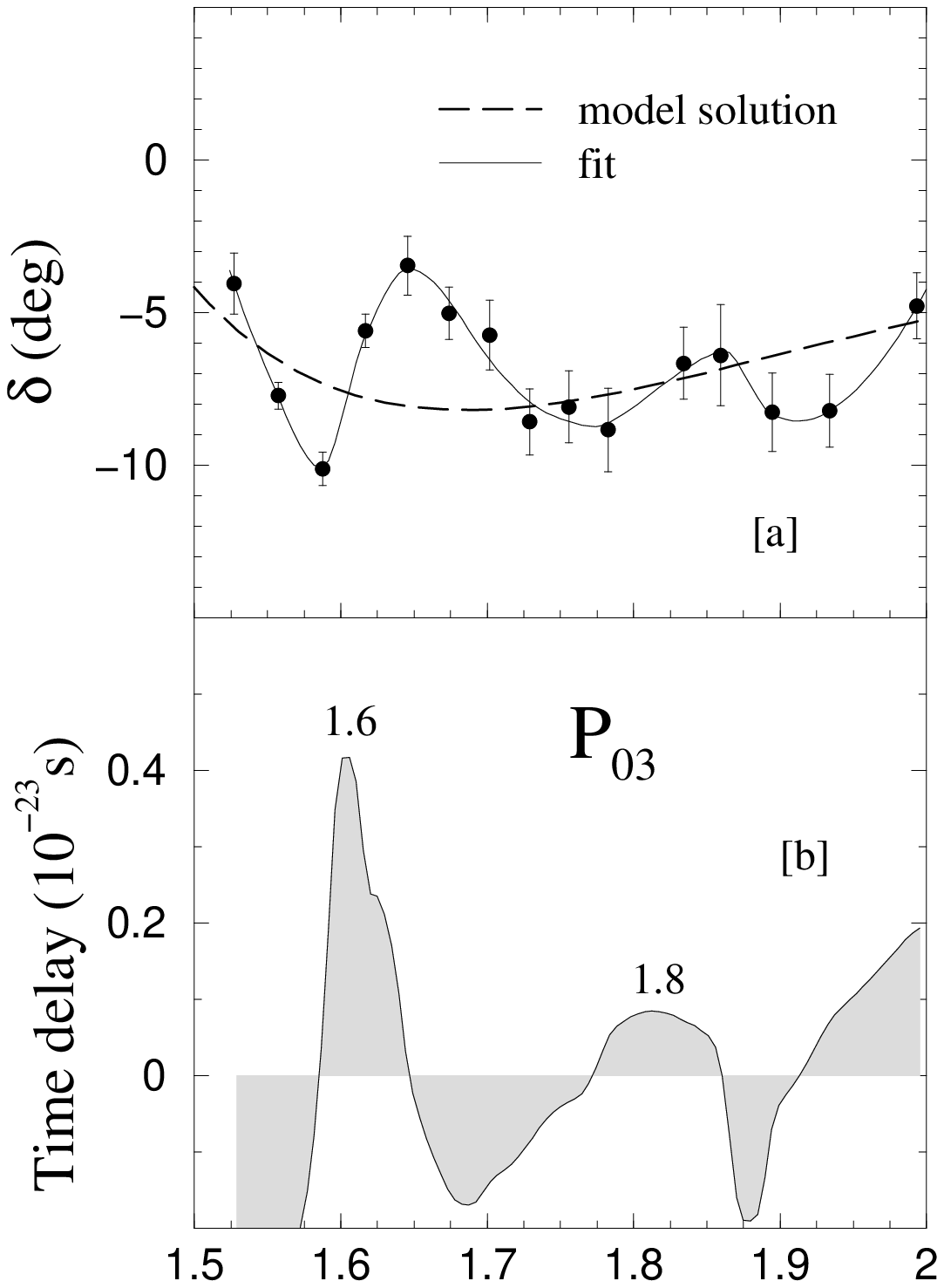}}
\caption{Same as Fig. 2, but for the $P_{03}$ 
partial wave.}
\end{figure}
Being motivated by our earlier experience with the meson and
unflavoured baryon resonances\cite{we3we4}, where small fluctuations in 
the single energy values of the phase shifts gave rise to time delay peaks
corresponding to lesser established resonances, we decided to perform
a time delay analysis of the single energy values of 
phase shifts in $K^+N$ elastic scattering 
too\cite{weKN}. In Figs. 2, 3 and 4 we show the time delay distributions
obtained from fits to the single energy values of the phase shifts.
It is interesting to note a peak at 1.545 GeV in the $D_{03}$ partial
wave which comes very close to the discovery of the $\Theta^+$ 
from recent cross section data. The peak at 1.64 GeV agrees with some
of the predictions\cite{jenake} of a $J^P = 3/2^+$, $D_{03}$ partner of
the $\Theta$(1540). In Figs. 3 and 4 we see that the resonances occur
at exactly the same positions, namely, 1.6 and 1.8 GeV in the case of
the $P_{01}$ and $P_{03}$ partial waves which are $J=1/2, 3/2$ partners. 
The $J=3/2$ partners of the $\Theta^+$ have also been 
predicted\cite{gloclo} to lie in the region from 1.4 to 1.7 GeV.
An interesting discussion on the possible spin and parity of the 
$\Theta^+$ by examining the two kaon photoproduction cross sections
can be found in Ref.[33].

In closing, we note that the three peaks, 
namely, 1.545 in the $D_{03}$ and 1.6 and 1.8 GeV 
in the $P_{01}$ and $P_{03}$ partial waves are in very good 
agreement with the experimental values\cite{azl}, 1.545$\pm$.012, 
1.612$\pm$.01 and 1.821$\pm$.11 GeV of the resonant structures in the 
$p K_s^0$ invariant mass spectrum. We can then identify the time delay 
peak in the $D_{03}$ partial wave to be the $\Theta^+$.  
We summarize the findings in literature along with the resonances found
using time delay in Table 2. 

\begin{table}[h]
\tbl{Determination of pentaquark resonances from different techniques in
literature}
{\begin{tabular}{@{}ccccc@{}}
\hline
\multicolumn{5}{c}{}\\[1ex]
Partial& Time delay& Time delay&Poles & Argand \\
wave  &(from fits) & (SP92)  && diagrams \\
\multicolumn{5}{c}{}\\[1ex]
\hline
&& &&\\
$S_{01}$ & 1.74 && 1.71\cite{roiesnol}& \\
 & 1.85 &1.85\cite{arndt}&& 1.85\cite{aaron}\\
&& & & \\
$P_{01}$ &  1.6 & 1.57&&\\
         & 1.81 & 1.83& 1.83\cite{arndt}& 
1.85\cite{aaron}\\
 & & &&1.78\cite{nakajima}\\
$P_{03}$ & 1.6 & & &\\
         &1.8 & && \\
 & & &&\\
$P_{13}$ &1.6& 1.5& & \\
 &1.75 & 1.75&1.81\cite{arndt} & \\
         & 1.92 & & & 1.93\cite{nakajima}\\
         &  & & & 1.9-2.0\cite{kato}\\
$D_{03}$ &1.545& 1.49&& \\
 &1.64& & & \\
         &1.77& 1.81& 1.79\cite{arndt}& 
1.85\cite{aaron}\\
 & &&& 1.91\cite{nakajima}\\
$D_{15}$ & 2.0 & 2.0& 2.1\cite{arndt}& \\
\hline
\end{tabular}}
\end{table}

{\it Note: The focus of this talk has been on the use of the time delay method 
for identifying pentaquark resonances. Hence we have not quoted all
possible references on pentaquarks/exotics which appeared recently}.

\end{document}